\documentclass{IEEEcsmag}

\usepackage[colorlinks,urlcolor=blue,linkcolor=blue,citecolor=blue]{hyperref}

\usepackage{upmath}

\newcommand{\rv}[1]{{\color{black} #1}}

\begin{document}


\title{Challenges and Opportunities for Beyond-5G Wireless Security}

\author{Eric~Ruzomberka}
\affil{Department of ECE, Princeton University}

\author{David~J.~Love}
\affil{Elmore Family School of ECE, Purdue University}

\author{Christopher~G.~Brinton}
\affil{Elmore Family School of ECE, Purdue University}

\author{Arpit~Gupta}
\affil{Department of CS, University of California, Santa Barbara}

\author{Chih-Chun~Wang}
\affil{Elmore Family School of ECE, Purdue University}

\author{H.~Vincent~Poor}
\affil{Department of ECE, Princeton University}


\begin{abstract}
The demand for broadband wireless access is driving research and standardization of 5G and beyond-5G wireless systems. In this paper, we aim to identify emerging security challenges for these wireless systems and pose multiple research areas to address these challenges.
\end{abstract}

\maketitle


\section{Introduction}
Over the past few decades, wireless communication has evolved from a luxury to a necessity for people all around the world.  Wireless access is now essential to the economy, educational system, and critical infrastructure.  Our dependence on wireless is projected to become even more pervasive, as both the number of wireless-enabled devices and demand for wireless access continue to grow roughly exponentially.  

To meet this widespread demand, industry and academia have responded with much research, standardization, and development of wireless broadband networks.  This has led, for example, to recent standards such as 4G LTE and 5G NR through 3GPP and various versions of IEEE 802.11.  Current physical-layer and network-layer research will have an impact on future releases of current standards and beyond-5G standards.  

Several technology themes have emerged in current research and development for wireless systems:\\
\textbf{More Antennas:}  Multiple-input multiple-output (MIMO) systems leverage multiple antennas at both the transmitter and receiver(s).  Intuitively, the multiple antennas allow the spatial degrees-of-freedom of the wireless channel to be used to construct multiple data streams at the same time and in the same band.  The potential throughput and reliability advantages increase  with the number of antennas.\\
\textbf{Non-Traditional Frequencies:} Most commercial wireless systems today operate in bands that are below 6 GHz. This limits the available bandwidth that can be allocated to different users. To alleviate this, there is a push to look at frequency bands that have been considered unusable for wireless broadband in the past.  In 5G NR, the millimeter wave bands (roughly defined as 20-100 GHz) are included in the standard, allowing substantially higher bandwidths.  Currently, there is much interest in communication systems that have carrier frequencies above 100 GHz as well.  \\
\textbf{Spectrum Sharing:} The demand for bandwidth in the sub-6 GHz frequency ranges is motivating regulators to consider new forms of licensing.  There are several emerging successful use cases where a primary network operates in a band and a secondary network is allowed to use the band as long as minimal interference is generated to the primary network.  In the future, networks may be able to dynamically share a frequency band in real-time.    \\
\textbf{Software-Defined Radios and Networks:}  Both radio-level and network-level functionality are moving to platforms that can be readily reconfigured through software.  The capabilities of software-defined radios (SDRs) continue to improve, and they are widely available commercially-off-the-shelf throughout the world.  Software-defined networks (SDNs) are also in widespread use due to their offered flexibility in network reconfiguration.  There is also the emerging open-radio access network (O-RAN) movement that may soon revolutionize how wireless networks are implemented and commercialized. 

These technology areas have led to many vibrant research directions in communications and networking.  At the same time, these areas have generated unforeseen security challenges that will greatly impact radio and network operation moving forward.  The vast majority of research \rv{at the physical layer has} focused on optimizing performance metrics such as spectral efficiency (bits/second/Hz), network efficiency (bits/second/Hz/area), and latency (seconds), while often ignoring any security constraints that arise.

Wireless communication and networking researchers can no longer afford to conduct research that does not consider security challenges.  In this paper, we discuss the emerging security challenges of beyond-5G wireless systems, and explore research directions aimed at addressing these challenges.  


\section{Evolution to 5G and Beyond} \label{sec:evolution}
\begin{figure*}[t]
    \centering
    \includegraphics[width=2\columnwidth]{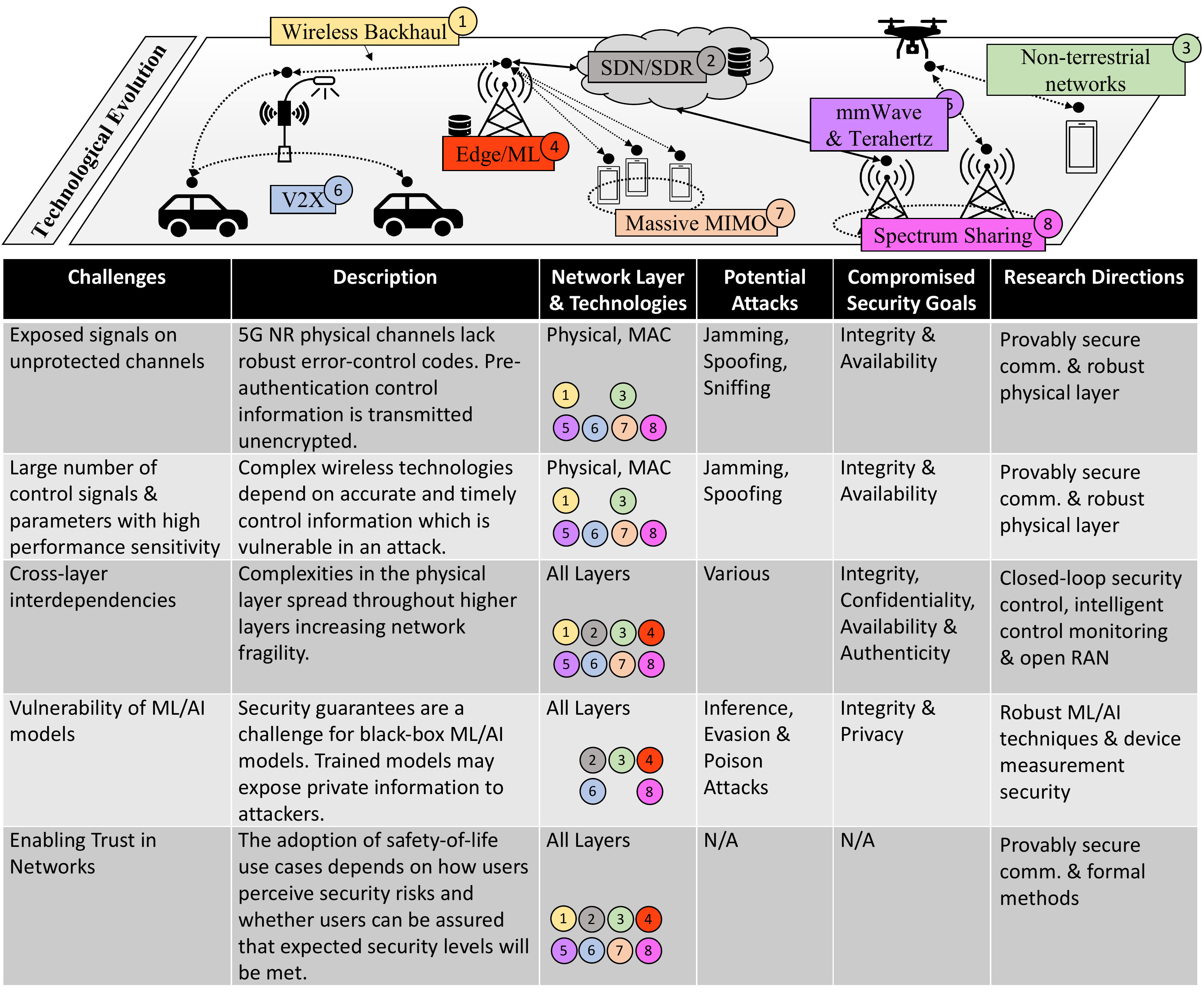}
    \caption{A high level overview of the security challenges faced by 5G and beyond-5G wireless networks and associated wireless technologies. Abbreviations: SDN -- software defined networking; SDR -- software defined radio; ML -- machine learning; mmWave -- millimeter wave communication; V2X -- vehicle to everything communication; MIMO -- multiple input multiple output.}
    \label{fig:NSpolicy}
  \end{figure*}

5G wireless systems are governed by the 5G NR standard produced by 3GPP. These systems, as well as beyond-5G communications systems, will include innovations at the physical layer as well as higher layers. We will briefly summarize several of these innovations depicted in Figure~\ref{fig:NSpolicy}:

\textbf{Massive Multiple-Input Multiple-Output (MIMO)}:  MIMO communication systems utilize multiple antennas at the transmitter and receiver to enable wireless signals to be processed spatially.  When employed at the base station, multiple antennas allow the transmit signal to be adapted spatially to changing channel conditions.  This spatial adaptation, usually accomplished through precoding, can yield a substantial capacity and/or reliability enhancements. Spatial resources can also be distributed across multiple users, which allows the downlink or uplink to support multiple users at the same time in the same frequency band.

The benefits of MIMO continue to scale as the number of antennas increases.  Generally speaking, each new standard release increases the maximum number of supported antennas.  We are now approaching massive MIMO systems, which use a much larger number of antennas at the base station.  Massive MIMO systems can yield large network spectral efficiency enhancements because in many cases the user channels approximately orthogonalize, which enhances the ability to support multiple users. \rv{In 6G, MIMO systems may be combined with novel technologies like reconfigurable intelligent surfaces (RIS) which enable programmable control of the wireless environment for further improvements in spectral efficiency and coverage.}

\textbf{Millimeter Wave and Terahertz}:  Before 5G, the majority of commercial cellular systems used bands located below 6 GHz.  In particular, frequency bands above 20 GHz received little to no commercial use. This changed with 5G NR deployments that include Frequency Range 2 (FR2) bands, which lie in the 25-30 GHz range. 

Academic and industrial research has referred to systems operating in the 20-100 GHz range as millimeter wave systems. Future releases of 3GPP for 5G are expected to utilize bands up to 100 GHz. Terahertz wireless communication, roughly defined as using the bands from 100-300 GHz, is a major research area for 6G wireless. Systems using higher frequency bands like millimeter wave and terahertz will generally require directional transmissions to focus energy at the receiver. Narrow-beam directional transmission can be very challenging because  tight alignment between the transmit and receive apertures typically requires careful measurement, sounding, and coordination.  Luckily, the channels encountered at these higher bands often have sparsity properties that can be leveraged to reduce alignment overhead time. \rv{Beyond the THz bands, researchers continue to explore visible light communication (VLC) as a promising candidate for 6G indoor wireless access as a low-cost access technology relative to THz access.}



\textbf{Wireless Backhaul Integration}:  Base stations in cellular systems are often connected to high-rate fiber links on the backhaul. These links carry the aggregated communication signals from users communicating with the backhaul to the wider-scale networks.
Unfortunately, establishing new wired backhaul connectivity is impractical in many settings due to cost or other factors.  To overcome this, many networks have utilized in-band wireless backhaul solutions.  The general idea is to allocate a portion of the network's spectrum for carrying backhaul signals between cells. The typical approach, however, has been to operate these backhaul networks in an unoptimized way that is uncoupled from the wireless links supporting user access within cells.

5G NR has moved to integrate backhaul links with access links. This becomes critical when networks are required to support high-reliability or low-latency traffic. Backhaul integration may also see wide use in rural deployments, where small user densities pose challenges to justifying wired backhaul costs.

\textbf{Spectrum Sharing and Unlicensed Access}:  Spectrum sharing has been a point of commercial and academic interest for several decades.  The general idea is that a band can be used by multiple networks or multiple radio systems at the same time.  Unlicensed access for cellular was introduced in LTE and has continued into 5G NR in Release 16.  

Spectrum sharing in a licensed band is usually formulated as a primary-secondary user sharing problem.  The primary network has preferred access to the band.  A secondary network can only use the band under the condition that it creates minimal impact on the operation of the primary network.  This form of sharing had notable use first in the TV whitespace bands, and more recently in the Citizens Broadband Radio Service (CBRS) band shared by the federal government (i.e., primary user) and the private sector (i.e., secondary user). Primary-secondary sharing is also being considered for mitigating interference between radar altimeters used in aviation and 5G.

Beyond-5G access could move to far more exotic spectrum sharing models. For example, collaborative and lightly regulated access was shown to be possible in the DARPA Spectrum Collaboration Challenge (SC2).  In this framework, an agreed-upon protocol is used to facilitate spectrum sharing between multi-node networks operating in the same band.  The end goal would be to have networks that come close to completely filling the time-frequency resource grid at every point in a geographic area subject to appropriate fairness constraints. \rv{As another example, sharing could incorporate backscatter communication in which devices passively reflect signals from their ambient wireless environment to communicate over short distances.}

\textbf{Non-Terrestrial Networks}:  The widespread deployment of wireless broadband is taken for granted in much of the developed world.  This is mainly because most users are congregated in urban areas, suburban areas, or along major roadways.  Even in highly-developed countries, access in lightly populated rural areas falls below most acceptable thresholds.  The problem is even further exacerbated in third-world countries.  

To approach ubiquitous connectivity without incurring unrealistic capital investments, operators have started to consider non-terrestrial networks (NTNs) through technologies such as unmanned aerial vehicles (UAVs), high-altitude aerial platforms, and low-earth orbit (LEO) satellites.  NTN commercialization is still in its infancy, but it is a point of major commercial, academic, and government interest.

\textbf{Machine Learning for Wireless}:  Machine learning has infiltrated almost all areas of engineering over the past decade \cite{Eldar2022}. Deployment of learning-based solutions in wireless networks lagged behind other domains (e.g., computer vision) for several years due to a lack of comprehensive and publicly available signal measurement datasets. Generating representative datasets in this setting is challenging because of the large number of environmental parameters and interference condition variations that impact wireless behavior. However, tabulated databases and open access emulators have begun to emerge in the past few years, from DARPA's RF Machine Learning Systems (RFMLS) program to the National Science Foundation's (NSF's) Platforms for Advanced Wireless Research (PAWR) initiative.

6G-and-beyond standards are envisioned to include several machine learning algorithms for network control optimization. One class of techniques under consideration is deep learning approaches to channel estimation, which have shown improvements over conventional linear estimation techniques. Another area of investigation is reinforcement learning approaches for controlling the actions of network elements, such as RIS.

\textbf{Vehicle-to-Everything (V2X) Communication}:  V2X communication has been discussed for several decades without much commercial activity.  Over the last few years, the push for autonomous vehicles and improved safety systems has motivated cellular systems to be designed for V2X operation, often referred to as cellular V2X (C-V2X). 

Cellular V2X has been furthered by the recent FCC decision on repurposing the 5.9 GHz band. Additionally, Release 16 of 5G NR includes sidelink enhancements that can be of use for V2X.  Academic and commercial research on V2X continues to grow, and it will evolve further as the V2X use cases grow.


\textbf{Software Defined Radio (SDR), Open-Radio Access Network (O-RAN), and Software Defined Networking (SDN)}:  Historically, radios and networks were implemented using hardware-centric techniques that limited their ability to be upgraded and adapted once deployed.  Increasingly, however, radios and networks are being implemented using software and more general purpose hardware.  This revolution arguably started with the development of SDRs for experimental use in academic and research environments.  Software-defined architectures then became popular in networking, with SDN techniques now allowing networks to be managed and controlled without hardware changes.  
For 5G and beyond-5G networks, there is much interest in O-RAN, which could open network blocks to users and remove much of the proprietary opaqueness around wireless operation.

\section{Security Challenges}\label{sect_security}
The innovations for 5G and beyond wireless all share a few common traits.  First, they each provide a level of time, frequency, and space \textit{adaptability} that far exceeds anything found in past commercial systems.  Second, 5G and beyond networks have a level of \textit{flexibility} far exceeding past implementations.  The flexibility and adaptability properties are coupled out of necessity since any ``optimal" network/radio operation will require ``control knobs" that can be turned to achieve this optimality.

At the same time, flexibility and adaptability combine to generate an attack surface for 5G and beyond systems that is orders of magnitude more challenging to secure than past wireless networks.  All layers of the network interact in a carefully coordinated way.  While in the past, most of this interaction was hard wired, today and in the future, an attacker can target any layer of the radio network and have a cascading impact on other layers. 

 \textbf{Standardization:} In most cases, the signals and inputs/outputs of the procedures that define the network operation are publicly available. Standards-compliant devices must follow these specifications, which has allowed the wireless industry to achieve a admirable level of interoperability between manufacturers and devices, while also exposing such information for potential attackers. \rv{The push towards O-RAN has increased the amount of information exposed to attackers via standardization.}
 
 \rv{At the same time, security has been a key focus of 5G standardization. At the higher network layers, this focus has resulted in many 3GPP security enhancements over 4G including improved authentication, privacy protections for subscriber identity, and integrity and confidentiality protections for data/control plane traffic. Further efforts have resulted in O-RAN security enhancements to recently opened RAN interfaces and functions. 
 
 At the physical layer, however, security has been less of a priority. As in 4G, 5G NR pre-authentication signals exchanged between the user and base station remain unprotected. Furthermore, the standard leaves many security challenges open, including challenges surrounding SDN, ML and AI. Currently, standards groups like the ETSI Industry Specification Group on Securing Artificial Intelligence (ISG SAI) are at work on some of the later challenges.}
 
\begin{figure}[t]
    \centering
    \includegraphics[width=\columnwidth]{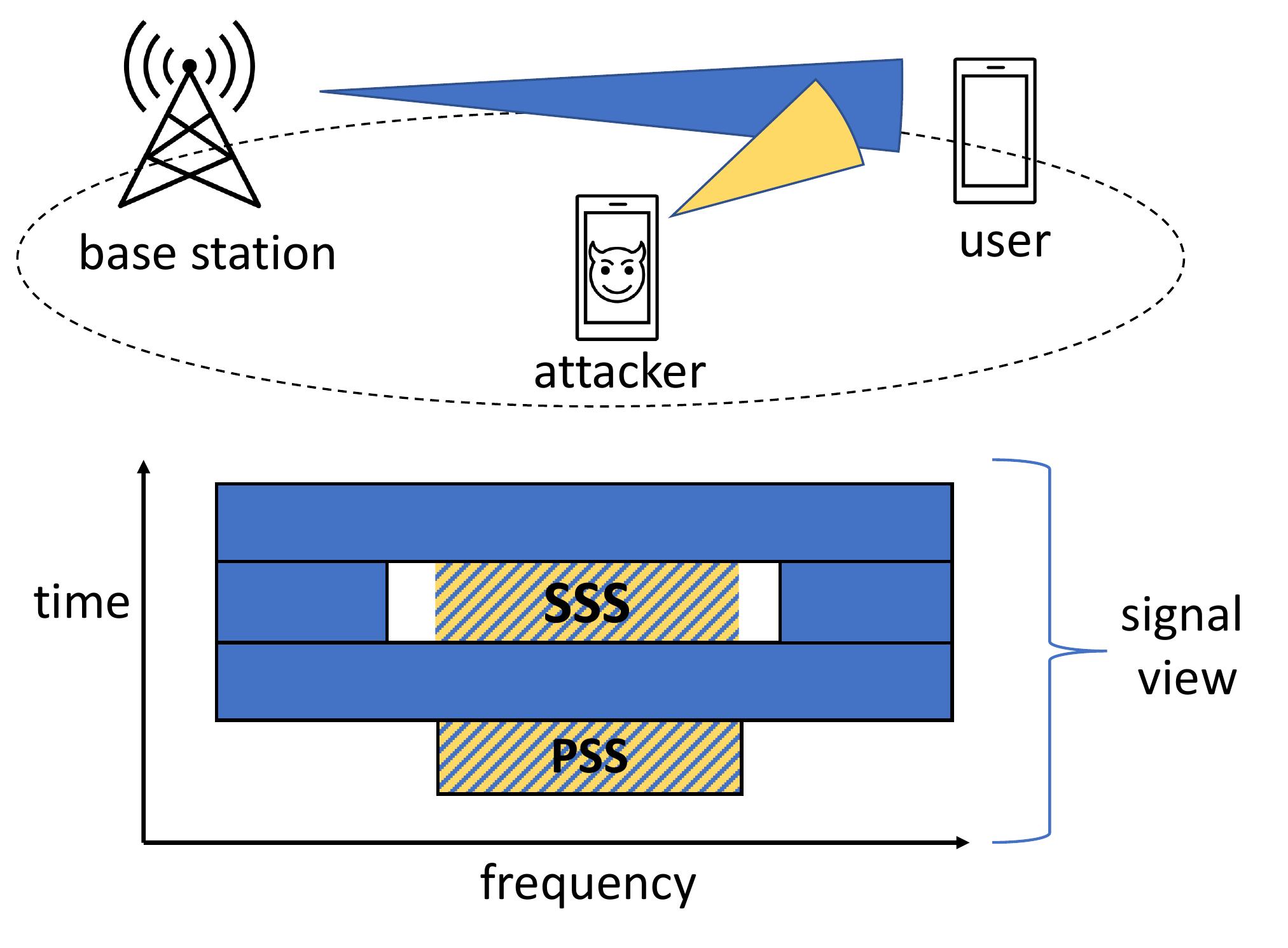}
    \caption{\rv{Physical layer signals exchanged between the base station and user are vulnerable to adversarial manipulation. For example, an attacker can easily prevent a user from synchronizing with a base station by manipulating the exposed Primary Synchronization Signal (PSS) and Secondary Synchronization Signal (SSS) located in the Synchronization Signal Block (SSB).}}
    \label{fig:jam}
  \end{figure}

\textbf{Exposed Physical Signals:} At the radio layer, procedures for timing, control, and channel estimation have all become extremely important for protocol operation and thus are a target for malicious manipulation. \rv{As depicted in Fig. \ref{fig:jam}, an attacker from outside the network can manipulate these procedures by injecting wireless signals over top of legitimate signals transmitted by base stations and users. Injected signals can corrupt legitimate signals and override the procedure's error-control mechanisms. Such attacks are increasingly feasible, made possible by accessible hardware in the form of low-cost commodity SDRs and access to publicly available standards. Currently, 5G physical channels lack robust error-control codes or cryptographic protections and thus are exposed to manipulation. These attacks can occur anywhere geographically in the network and compromise the availability or integrity of physical layer communication.}

For example, an attacker could target the synchronization signals of a base station and the random access signals from a base station, disrupting the ability of users to reliably synchronize in time and frequency. In 5G NR, these signals are transmitted in the synchronization signal block (SSB) and physical random access channel (PRACH), respectively. Because of the reliance on time-frequency resources, even relatively slight timing offsets can reduce the effective communication rate to zero. For example, jamming or spoofing the synchronization signals can obstruct the ability to get accurate timing and, thus, stop communication altogether. 

The number of wireless control signals and control parameters continues to grow with each new standard release. \rv{This trend is driven by increasingly complex RAN technologies such as massive MIMO, millimeter wave \& terahertz communication, wireless backhaul integration, non-terrestrial networks, spectrum sharing, and V2X, all which heavily rely on precisely set control parameters for efficient operation.} Both the uplink and the downlink carry significant amounts of diverse control information.

Control signals carry critical details such as scheduling requests, resource allocations, channel state information (CSI) feedback, ACK/NACK signals, and power control information.
Attacking one or more of these control signals could potentially eliminate the ability of a base station to successfully communicate with its users. \rv{Such attacks can be mounted stealthily with very little transmit power from an attacker's radio due to the lack of robust error-control codes in the 5G NR control channel \cite{Ludant2021}.}

\textbf{CSI Vulnerability:} Adaptability in modern and future communication systems is achieved based on information about the current propagation conditions of the wireless signals \cite{Love2008}. These propagation conditions are often encapsulated in an input-output model that uses a linear-time varying filter, commonly referred to as a channel model. This channel is usually represented as one or more time-varying vectors or matrices.  Typically, the time variation is assumed to be sufficiently slow relative to the data symbol rate to allow the channel to be modeled as constant for some number of discrete time indices.

At any point in time, nodes within a wireless network aim to acquire channel state information (CSI) for these models.  When the node is a receiver, information about the channel that the received signal experienced can be acquired by having the transmitter send a known signal, usually referred to as a pilot or reference signal.  The receiver will then use this known signal to estimate the channel conditions.  When the node is a transmitter, the aforementioned channel conditions can be attained using feedback, which is especially useful in frequency division duplexing (FDD), or inferred from a receive channel using reciprocity, which is primarily only possible in time division duplexing (TDD). This channel information is subsequently used to allocate precious resources in space, time, and/or frequency. Decisions on signal parameters such as modulation order, rate allocation, power level, precoding, and scheduling are made using this CSI. \rv{Certain radio technologies have particularly high performance sensitivity to CSI accuracy, including technologies that require narrow transmit and receive beams (e.g., massive MIMO, millimeter wave \& terahertz communication) or stringent performance requirements in highly mobile environments (e.g., non-terrestrial networks, V2X).}

As a result, when this CSI is corrupted or unreliable, there can be a large loss in network spectral efficiency.  If the reference signal or received signal during training are corrupted, CSI will be unreliable, which will cause all decisions made based on this CSI to be unreliable.  In other words, the CSI acquisition process creates significant vulnerability for the entire network.  Even if channel estimation is done properly, CSI feedback is still vulnerable: this feedback and its associated control signals have to be carefully protected.

\textbf{Cross-layer Inter-dependencies:} In summary, modern wireless standards have grown increasingly fragile, beginning with protocol designs at the physical layer. Further, the goal of optimizing wireless networks over large geographic areas has led to network-wide transmitters and receivers, with resource allocation decisions shifting from a local view to a network view. 
The push to increase date rates across multiple users operating across multiple access points or base stations has caused networks to be architected with complicated inter-dependencies spread across all layers of the protocol stack \cite{jen-nick:sdn18}.

\rv{As a result of these complicated inter-dependencies, today's networks have become difficult to build and operate with strict security goals in mind. Small changes to one layer in one locality of the network can cause unintended and unspecified changes in other layers throughout the global network. In turn, security vulnerabilities can be tricky to identify and resolve. Simply operating these fragile networks is a labor-intensive task, requiring input from a team of experts. Thus, the process for securing today's network is a slow. In future networks, these challenges will compound as networks scale to service more devices with more heterogeneous security requirements.}

\rv{\textbf{ML/AI Vulnerability:} ML and AI techniques are likely to play a key role in the security framework used in next generation wireless networks. Programmable intelligence capabilities will be supported by edge devices and RAN controllers, where the latter are introduced via the O-RAN SDN architecture. From a security point-of-view, ML/AI techniques can add robustness back into fragile networks by automating many security tasks, including network monitoring, threat detection and policy recommendations.

ML/AI can, however, be a double edged sword for network security. Many models depend on sophisticated black-box algorithms which in turn further enlarge the overall attack surface of the network. Thus, ML/AI can become the target of an attack. In particular, ML/AI is susceptible to adversarial interference during model training whereby an attacker contaminates (i.e., poisons) training data sets via either modification of true data or injection of false data. These so called ``poisoning'' attacks can compromise the integrity of the model by degrading performance (e.g., classification accuracy) or by introducing a backdoor that can be exploited after training. An attacker may have various levels of knowledge and access pertaining to the model and training data.  

In another instance, an attacker can exploit vulnerabilities in a trained model. This can occur in an evasion attack in which the attacker chooses inputs to induce erroneous behavior in the model, or in an inference attack in which information related to the training data is extracted from the model. Evasion attacks compromise the integrity of the model while inference attacks can compromise user privacy. Again, the attacker may have various levels of knowledge about the model and input data. At the physical layer, effects from the wireless channel like channel fading can have a limiting effect on this knowledge, e.g., by causing the model and the adversary to observe different features in the input data \cite{sagduyu2019adversarial}. All three vulnerabilities must be addressed prior to giving ML/AI a prominent role in a network security framework.

As discussed earlier, ML and AI will be key enablers for network management and optimization. Their application is likely to extend to most network domains, including signal processing at the physical layer, routing at the MAC layer, spectrum sharing solutions and intelligence-critical use cases like V2X. Thus, the above ML/AI vulnerabilities are best viewed as a security concern that touch all parts of network.}

 
\rv{\textbf{Trust in Networks:} Wireless networks are supporting new safety-of-life use cases with each new generation. Autonomous driving and road safety use cases in V2X are examples in which compromised network security can result in serious harm to human safety and property. It is clear that the global adoption of these use cases depends on how users perceive security risks and whether users can be assured that the network will meet expected security levels. To enable trust, networks must be able to verify to users that security measures meet expected levels.}

\textbf{Interdisciplinary Research:} Addressing the beyond-5G security challenges is further confounded by the fact that the associated problems have traditionally been spread over different research communities. Physical layer signal processing is more commonly associated with the communication theory and information theory communities that focus on analyzing the fundamental limits of transmission, such as achievable data rates and reliability, and then developing schemes to approach these fundamental limits.  Security is typically viewed as something left to higher layers, with recent learning-based approaches responding to increasing adversarial sophistication.  The softwarization of communication networks has emerged from computer science and computer engineering, which focus more on an algorithmic approach to experimentation.

\section{Research Directions}
There is an urgent need for research that addresses this myriad of emerging security challenges for 5G and beyond networks. We will briefly highlight several important areas for investigation.

\subsection{A New View of Physical Layer Security}
As mentioned earlier, physical layer communications have been an area of great importance to the information theory, communication theory, and signal processing research communities. This has generated a large body of work centered on i) understanding point-to-point and, later, multipoint-to-multipoint wireless link capacities; and ii) developing new signaling approaches, modulation techniques, codes, waveforms, and other methods to approach these capacity limits. 

The focus of these works has typically been a rigorous engineering approach where wireless channels are represented using mathematical models. The benefits of this area of research are undeniable. Techniques driving current standards such as MIMO wireless communications, orthogonal frequency division multiplexing (OFDM), adaptive modulation and precoding, capacity-approaching codes (e.g., turbo, low-density parity check, and polar codes), and signal processing for new bands (e.g., millimeter wave) are all outputs of this communication engineering approach.  

Security, however, has not been a typical constraint enforced in physical layer research.  When most researchers in communications and information theory think of physical layer security, they think of problems related to the wiretap channel \cite{Wyner1975}.  In this setup, a transmitter wants to convey information to a receiver, but the transmitter's signal can be overheard by an eavesdropper node.  The goal is to understand the fundamental limits on what communication rate can be achieved between the desired transmitter and receiver subject to constraints on how much information can be conveyed to the eavesdropper. There is a large body of work in this area, and there are many sophisticated network configurations that generalize this eavesdropper problem.

Today's security challenges, however, extend far beyond the eavesdropper setting.  The challenges outlined above necessitate a much different approach to communications and information theory.  Luckily, however, there are many valuable techniques from these areas that can be built upon to extend the frontier of security research.  
The typical approach of characterizing fundamental limits, and then developing practical schemes to achieve these limits, could yield new innovations for commercialization. We discuss two particular research directions/objectives in the following.

\textbf{Provably Secure Communications:} Most security research has been conducted using an ``arms race" approach.  New adversarial attacks are developed and/or observed, and techniques to combat these attacks are then researched. A provably secure approach to communications would instead develop mathematical models for communication or network operation that account for adversarial attacks in the first place.  These models would then be used to characterize the capacity of adversarial channels by deriving the tightest possible bound on the communication rate subject to \textit{any} adversarial attack, and then finding schemes that can achieve rates of communication up to this bound in the limit as the encoding block size grows large.  

At first glance, this type of research might seem daunting and potentially inapplicable to reality.  However, there are a variety of cases where this adversarial capacity can be fully analyzed and employed in practice. For example, consider the case of binary communication between a transmitter and receiver in the presence of an adversary.  To enforce a practical model on the operation of the adversary, we could assume that the adversary can flip or erase the bits received up to some fractional bound. Surprisingly, it has been shown in the literature that the capacity of such a communication model can be fully characterized for different cases of adversarial knowledge of the message \cite{Suresh2022}. Therefore, no matter what strategy the adversary takes, it is always possible to find a scheme that achieves any rate up to the capacity in the limit. 

The binary channel is applicable to many wireless communications scenarios, but there is a need to understand the adversarial capacity for many of the more general wireless input-output models as well. In many cases, especially those when the adversary has uncertainty of the legitimate channel, it can be shown that the adversary's best strategy is to simply inject noise \cite{Kashyap2005}. 
However, these results do not necessarily generalize to all relevant communication models, which is a question that requires significant research attention \cite{Chorti2022}.

\textbf{Robust Physical Layers:} Today's physical layer innovations achieve performance levels on various key metrics that were unfathomable in past wireless generations.  However, the dependence on channel information and adaptation renders these networks susceptible to adversarial behavior.  To harden physical layer operation, several new research topics must be addressed.

For one, channel training and channel estimation must be reevaluated.  An adversary could employ various attacks to remove the ability to accurately estimate the wireless channel depending on the adversary's knowledge of the training sequence and timing operation.  There is a critical need to formulate and experimentally verify signal models that capture such scenarios.  Using these models, we can then proceed to characterize how to handle these attacks.  In doing so, it is important to consider that most practical adversarial setups will not have an infinitely powerful adversary.  The adversary will almost always be limited in its ability to corrupt, compute, or receive information.  Accurate adversarial models will accurately encapsulate these limitations and lead to new training schemes.


Accurate CSI acquisition is only part of the problem.  Communication schemes, many of which adapt to this CSI, are usually designed for input-output models with fading channels and additive noise.  As mentioned before, CSI corruption and interference that does not follow additive noise are usually ignored.  Research is thus needed on communication schemes that can confront CSI and interference issues and attain a provable level of performance in the presence of security threats.

\subsection{Programmable Networks} \label{formal}
Many of the above challenges are closely related to the programmable capabilities of modern networks. In particular, network properties like flexibility and adaptability result from the increasing ability of operators to directly program network elements in real time. This softwarization of networks is resulting in large attack surfaces, variable network behavior, and complex protocols and equipment which can be exploited by adversaries.
 

\textbf{Closing the Security Control Loop:}
Beyond 5G networks will require unprecedented visibility into and control over network behavior to handle their growing security challenges. Such capabilities are needed to quickly identify and respond to evolving threats. Today's mobile networks, however, are not yet able to deliver these security capabilities. This is in part due to the proprietary status of RAN components which are complex, opaque, and difficult to update. 
 
Meanwhile, 5G has begun to adopt concepts like disaggregation and openness for customization and control of network services. 
With disaggregation, a large complex bundle of network elements is broken up into easier-to-manage pieces. These pieces are often built from commodity hardware that can be directly programmed. In the next generation, disaggregation can enable networks to react quickly to evolving threats.

\begin{figure}[t]
    \centering
    \includegraphics[width=\columnwidth]{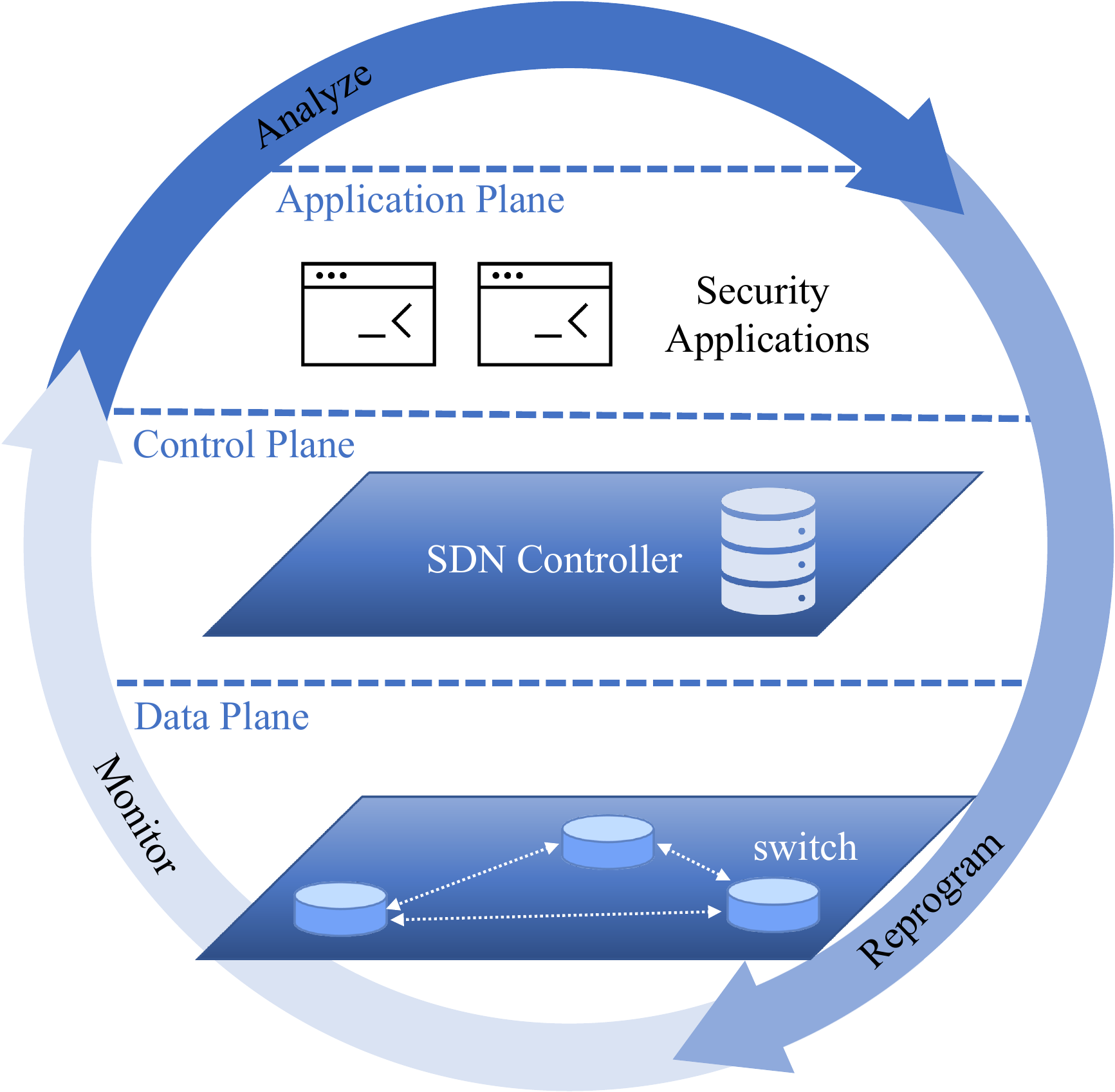}
    \caption{The security control loop. Network state data is collected by devices in the data plane, aggregated at the control plane, and exposed to security applications sitting above the control plane. In turn, these applications can analyze the data and recommend policy changes to the SDN controller. Policy changes are implemented by the SDN controller by reprogramming the data plane.}
    \label{fig:control_loop}
  \end{figure}

To illustrate this idea, we can consider the role of disaggregation in SDN. The SDN architecture, illustrated in Fig. \ref{fig:control_loop}, disaggregates the control and data planes, allowing an SDN controller to be placed logically above the data plane for centralized control of commodity switches. As a result of this placement, the controller is strategically positioned for visibility of all network flows. This visibility can be shared with security applications sitting above the controller, which in turn can request the controller to make security-related updates in the data plane. Hence, SDN security applications are well equipped to monitor, analyze, and reprogram the network.
 
In contrast to the above example, networks today monitor and update separately. Updating the network configuration in reaction to security events is made by an open-loop system uses little to no real-time feedback from network monitoring and requires significant human intervention. Hence, security update control is slow to respond to new threats. To address the stated challenges, beyond 5G networks must close the security control loop. The idea of closed-loop network management (often referred to as the ``self-driving network'') has gained a lot of traction in the networking community~\cite{jen-nick:sdn18,srn-report19}.
Research into self-driving networks is currently in the early stages, and efforts are needed to address the challenges unique to realizing a self-driving RAN. Compared to the core network, a RAN requires more comprehensive monitoring and control due to fast-changing channel conditions and limited bandwidth availability. 

A self-driving network for 5G RAN entails the following tasks: (i)~detailed and continuous monitoring of the state of the network, (ii)~making accurate inference decisions in (near) real-time about a myriad of different security events, and (iii)~effective and fast mitigation of each of these events. Realizing these tasks for 5G's RAN entails processing TBs of highly diverse and semantically rich data every second, which in effect rules out the use of off-the-shelf network telemetry tools and network streaming analytics systems~\cite{sonata-paper,univmon}.
Thus, closing the security loop for the 5G network requires developing an intelligent and resilient network telemetry stack. This stack should make the best use of limited and heterogeneous data plane resources to realize multiple concurrent network analytics tasks, including tasks requiring learning-based inference decisions. 

Developing such a telemetry stack requires interdisciplinary research efforts requiring networking, machine learning, and computer architecture expertise. We envision a ``layered" design, with a middle layer interfacing with a top layer that focuses on ``AI/ML for networking" and a bottom layer concerned with ``networking for AI/ML". The top layer will require the development of reinforcement learning (RL) frameworks responsible for coordinating the execution of a given set of concurrent analytics tasks in ways that best use a given set of hardware targets. The middle layer will serve as ``glue" between the top and bottom layers of the proposed telemetry stack with new learning-based algorithms specifically designed to determine effective query plans that ensure robustness to changes in input traffic and query workload. The bottom layer will require the development of custom hardware targets that are architected specifically with network analytics in mind---enabling flexible analytics tasks dynamically at scale.

\textbf{Open RAN:} Comprehensive monitoring and control has been a challenge for RAN operators. In principle, operators should readily be able to collect and aggregate security-relevant data from components across the RAN. Such data should inform the operator on how to tune the right ``control knobs'' and drive the network to a secure state. In practice, monitoring and control is limited by the proprietary status of many components and their interfaces.

Recent standardization efforts like O-RAN 
have begun to improve monitoring and control by allowing interoperability of RAN components made by different vendors. Interoperability is enabled by opening a number of interfaces between RAN components that were previously vendor-specific, which exposes network state information formerly hidden inside components. For security research, this exposure presents an opportunity for more comprehensive network monitoring and control.

By opening interfaces, we can expect that an O-RAN network will generate ever more security-relevant data from across the network. Combined with the programmability of the network, each switch, gate, controller, and access point can be viewed as a sensor capable of accurately measuring the network state in its locality. Each sensor can be directly programmed via its commodity hardware to report state information back to the operator. Hence, the operator can dynamically tailor each report to contain data as specific and fine-grained as needed. 

Lastly, there are still many open research opportunities for securing the O-RAN architecture itself. There are ongoing concerns that the combining of components from different vendors can lead to security issues. Also, open interfaces present a new attack surface that must be secured. Comprehensive research efforts are needed to address these issues.
 
\textbf{A Formal Approach to Mobile Security:} 
Today, most wireless communications research is focused on adding network functionality and increasing performance. Security is often treated as a second priority behind these two primary objectives. The challenges at hand require a new approach to wireless networking research where functionality/performance is optimized subject to security guarantees.  

Modeling such guarantees requires a formal approach to security. Formal methods will adopt a mathematical framework which allows a protocol designer to precisely describe security goals in the form of formal specifications. Given a mathematical framework, software tools can be used to systematically prove protocol properties and verify that the goals have been met. Most importantly, these tools will allow this to be done without relying on heuristic or ad hoc approaches that can lead to unexpected specification violations. 

Decades of research in networking have produced a number of mature formal tools that are ready to use, yet the application of these tools for wireless security faces a number of challenges. One major challenge is the immense difficulty in extracting formal specifications from the mobile standards, as 5G standards contain informal specifications that are written informally and thus may have different interpretations to different developers. Hence, before formal tools can be applied, these ambiguities must be resolved. Resolving these ambiguities, however, requires a large amount of effort from researchers with a wide-reaching expertise in formal methods, the specifications, and implementation.
 
The value-add from such legwork in formalization will be substantial, as demonstrated by recent standards contributions. Following the release of 5G security specifications in 2018, a small, but growing, body of research has extracted formal specifications for the analysis of 5G security protocols. A few important successes have come out of this work. For instance, vulnerabilities were discovered in the non-access stratum (NAS) and radio resource control (RRC) layer procedures \cite{Hussain2019}. 
These vulnerabilities were inherited from 4G and thus have laid dormant in production deployments for years.

\rv{The above example illustrates the potential for formal methods to foster trust in wireless security protocols. While the 5G standard is now complete, work will soon begin on the 6G standard, raising opportunities to incorporate formal security requirements and analysis during the pre-release phase of 6G standardization. In particular, 3GPP can better work with the research community to catch security vulnerabilities before the standard is released. Such opportunities can only be realized if closer relationships are fostered between standardization bodies and formal methods researchers.}


Lastly, before formal methods can be integrated into the standards, the community must answer an important question: \textit{how formal should the standards be?} While more formalization allows for stronger security guarantees, formal specifications can also hinder innovation. 
In the past, gaps in the standard have been a rich space for innovation, allowing manufacturers and operators to compete by offering better and differentiated products. It is up to the networking community to decide how to balance the above tradeoff between the innovation offered by the current standardization process and security guarantees offered by more formalized specifications. 

\subsection{Learning-based Innovations for Wireless Security} \label{MLsec}
The security challenges also motivate consideration for the plethora of data being generated and shared over today's wireless networks. At a high level, we can divide these measurements into two categories: those related to control operations of the network (e.g., for CSI acquisition), and those captured by user applications for intelligence building (e.g., location and camera information on a social media app). Both of these data types present challenges and opportunities for wireless security, and motivate machine learning approaches for enhancing beyond-5G threat resiliency.

\textbf{Intelligent Control Monitoring:} As discussed above, virtualization and standardization efforts such as O-RAN will allow nodes to collect and report network state information. These measurements are potentially useful from a security point-of-view, with programmable monitoring of the RAN playing an important part of the consolidated security control loop in Figure~\ref{fig:control_loop}. However, data collected from open interfaces is currently unorganized and has undefined value in a security context. Furthermore, we currently lack a clear road map to integrate this data into a security analytics framework. Moving forward, researchers must better understand how open data fits into this analytics road map.

In particular, as part of the O-RAN push, new intelligent tools will soon become available for integration into the security control loop. The proliferation of machine learning (ML) and artificial intelligence (AI) processing capabilities at the network edge presents an opportunity to identify adversarial behavior and mitigate threats based on rapid processing of local data. At a longer timescale, AI/ML and more general data science techniques can provide information on decisions made, recommend policy changes, and furnish other diagnostic information to human network operators.

Thus, programmable intelligence is likely to play an important role in closing the security update control loop. Still, significant research is needed to understand how existing AI/ML approaches must be tailored and/or augmented to provide robustness in the wireless security context. In this regard, one recent line of work, e.g.,~\cite{sagduyu2019adversarial,sahay2021deep} has examined learning-based approaches to signal detection and classification tasks in the presence of adversaries aiming to disrupt model training. A key takeaway from such efforts thus far has been that neural network methodologies can build resiliency against different types of attacks when they are combined with domain-specific feature engineering techniques -- such as crafting both time and frequency domain representations of the received signal -- to feed into the model.

In the wireless edge context, additional consideration must be given to the complexity of the AI/ML models being deployed for network security. In practice, the models must be executed rapidly based on new sensor measurements for detection of and response to emerging threats. They must also be updated/re-trained as information about new types of attacks becomes available. This becomes challenging when considering the cost of implementing contemporary deep learning models, consisting of potentially millions of model parameters, on resource-constrained computing nodes. On the other hand, a collection of lower complexity models (e.g., linear regression, support vector machines) deployed across edge nodes, each updated based on their own local sensor information with periodic parameter sharing would be much more scalable. Further research is needed to consider the tradeoff between complexity, accuracy, and robustness metrics for different network topologies under various attack models.

\textbf{Device Measurement Security:} Several mobile applications today, e.g., social media engines and augmented/virtual reality (AR/VR) platforms, are driven by AI/ML models trained on user-generated data. Traditionally, the building of models for such applications has occurred by transferring user data over the RAN to a datacenter for centralized training. However, transferring user data over over-the-air presents natural privacy/security risks, which are heightened by the expanded attack surface from open interfaces. The result is that users are less willing to share their data, which presents fundamental challenges to improving mobile intelligence applications.

In the last few years, federated learning has received significant attention~\cite{hosseinalipour2020federated}
as a technique for distributing model training across mobile devices. In the standard federated learning architecture, devices update models locally based on their own datasets, and their models are periodically aggregated and synchronized by a server. One of the attractive features of this framework from a privacy standpoint is that only models, rather than raw measurements, need to be transferred over the network. However, in certain instances, intelligent adversaries are still capable of recovering sensitive user information from models, e.g., through inference attacks on model parameters to reverse engineer training data. Additionally, adversaries can poison these models over-the-air through malicious measurement injection if they have model structure information.

Securing distributed model training is emerging as an important direction of research for next generation wireless networks~\cite{Eldar2022}. This will require efforts at the intersection of adversarial ML, which is typically from an application-layer perspective, and over-the-air threat modeling. Such techniques can draw from results in transmission coding and compression techniques at the physical layer, which will provide an added layer of security if they are injected into the design of models being shared over-the-air. Security guarantees produced by this work will make users less reluctant to participate in model training processes, in turn improving intelligence capabilities.

\section{Conclusion}

5G-and-beyond networks will help the world meet the ever-growing demand for wireless connectivity. These networks are enabled by recent and ongoing innovations in wireless networking technologies, many of which introduce a number of pressing security challenges. In this article, we have identified several primary challenges which remain largely unresolved despite current efforts from the communications and networking communities. 

To address these challenges, new approaches to wireless security research are needed. These new approaches span across research discipline (e.g., physical layer techniques, programmable networks and machine learning) and also across network layers. Foremost, these new approaches must prioritize security in the development, standardization, and deployment of the next generation of wireless networks.

\section{Acknowledgments}

This paper was supported in parts by ONR grants N00014-22-1-2305 and N00014-21-1-2472, and by NSF Grants CCF-1816013, CCF-2008527, CNS-2107363, CNS-2146171, CNS-2030299, EEC-1941529 and ITE-2226447.

\bibliographystyle{IEEEtran}

\bibliography{refs.bib}

\begin{thebibliography}{10}
\providecommand{\url}[1]{#1}
\csname url@samestyle\endcsname
\providecommand{\newblock}{\relax}
\providecommand{\bibinfo}[2]{#2}
\providecommand{\BIBentrySTDinterwordspacing}{\spaceskip=0pt\relax}
\providecommand{\BIBentryALTinterwordstretchfactor}{4}
\providecommand{\BIBentryALTinterwordspacing}{\spaceskip=\fontdimen2\font plus
\BIBentryALTinterwordstretchfactor\fontdimen3\font minus
  \fontdimen4\font\relax}
\providecommand{\BIBforeignlanguage}[2]{{%
\expandafter\ifx\csname l@#1\endcsname\relax
\typeout{** WARNING: IEEEtran.bst: No hyphenation pattern has been}%
\typeout{** loaded for the language `#1'. Using the pattern for}%
\typeout{** the default language instead.}%
\else
\language=\csname l@#1\endcsname
\fi
#2}}
\providecommand{\BIBdecl}{\relax}
\BIBdecl

\bibitem{Eldar2022}
Y.~Eldar, A.~Goldsmith, D.~G\"{u}nd\"{u}z, and H.~V. Poor, \emph{Machine
  Learning and Wireless Communication}.\hskip 1em plus 0.5em minus 0.4em\relax
  Cambridge University Press, 2022.

\bibitem{Ludant2021}
N.~Ludant and G.~Noubir, ``{SigUnder: a stealthy 5G low power attack and
  defenses},'' in \emph{Proceedings of the ACM WiSec}, Abu Dhabi, UAE, June 28
  - July 2 2021, p. 250–260.

\bibitem{Love2008}
D.~J. Love, R.~W. Heath, V.~K.~N. Lau, D.~Gesbert, B.~D. Rao, and M.~Andrews,
  ``An overview of limited feedback in wireless communication systems,''
  \emph{IEEE Journal on selected areas in Communications}, vol.~26, no.~8, pp.
  1341--1365, 2008.

\bibitem{jen-nick:sdn18}
N.~Feamster and J.~Rexford, ``Why (and {How}) {Networks} {Should} {Run}
  {Themselves},'' CoRR abs/1710.11583, 2017.

\bibitem{sagduyu2019adversarial}
Y.~E. Sagduyu, Y.~Shi, and T.~Erpek, ``Adversarial deep learning for
  over-the-air spectrum poisoning attacks,'' \emph{IEEE Transactions on Mobile
  Computing}, vol.~20, no.~2, pp. 306--319, 2019.

\bibitem{Wyner1975}
A.~Wyner, ``The wire-tap channel,'' \emph{The Bell System Technical Journal},
  vol.~54, no.~8, pp. 1355 -- 1387, 1975.

\bibitem{Suresh2022}
V.~Suresh, E.~Ruzomberka, C.-C. Wang, and D.~Love, ``Causal adversarial
  channels with feedback snooping,'' \emph{IEEE Journal on Selected Areas in
  Information Theory}, vol.~3, no.~1, pp. 69--84, 2022.

\bibitem{Kashyap2005}
A.~Kashyap, T.~Basar, and R.~Srikant, ``Correlated jamming on {MIMO Gaussian}
  fading channels,'' \emph{IEEE Transactions on Information Theory}, vol.~50,
  no.~9, pp. 2119 -- 2123, 2005.

\bibitem{Chorti2022}
A.~Chorti, A.~N. Barreto, S.~K\"{o}psell, M.~Zoli, M.~Chafii, P.~Sehier,
  G.~Fettweis, and H.~V. Poor, ``{Context aware security for 6G wireless: the
  role of physical layer security},'' \emph{IEEE Communications Standards
  Magazine}, vol.~6, no.~1, pp. 102--108, March 2022.

\bibitem{srn-report19}
N.~Feamster, A.~Gupta, and J.~Rexford, ``{Workshop on Measurements for
  Self-Driving Networks-Report},''
  \url{https://sites.cs.ucsb.edu/~arpitgupta/pdfs/measure\_selfdn\_workshop.pdf}.

\bibitem{sonata-paper}
A.~Gupta, R.~Harrison, M.~Canini, N.~Feamster, J.~Rexford, and W.~Willinger,
  ``Sonata: Query-driven streaming network telemetry,'' in \emph{Proceedings of
  the ACM SIGCOMM}.\hskip 1em plus 0.5em minus 0.4em\relax ACM, 2018, pp.
  357--371.

\bibitem{univmon}
Z.~Liu, A.~Manousis, G.~Vorsanger, V.~Sekar, and V.~Braverman, ``One sketch to
  rule them all: Rethinking network flow monitoring with {UnivMon},'' in
  \emph{ACM SIGCOMM}, 2016.

\bibitem{Hussain2019}
S.~Hussain, M.~Echeverria, I.~Karim, O.~Chowdhury, and E.~Bertino,
  ``{5GR}easoner: A property-directed security and privacy analysis framework
  for {5G} cellular network protocol,'' in \emph{Proceedings of the ACM CCS},
  London, UK, 11-15 Nov. 2019.

\bibitem{sahay2021deep}
R.~Sahay, C.~G. Brinton, and D.~J. Love, ``A deep ensemble-based wireless
  receiver architecture for mitigating adversarial attacks in automatic
  modulation classification,'' \emph{IEEE Transactions on Cognitive
  Communications and Networking}, 2021.

\bibitem{hosseinalipour2020federated}
S.~Hosseinalipour, C.~G. Brinton, V.~Aggarwal, H.~Dai, and M.~Chiang, ``From
  federated to fog learning: Distributed machine learning over heterogeneous
  wireless networks,'' \emph{IEEE Communications Magazine}, vol.~58, no.~12,
  pp. 41--47, 2020.

\end{thebibliography}

\end{document}